\documentclass{article}

\PassOptionsToPackage{numbers, compress}{natbib}
\usepackage[preprint]{neurips_2025}

\usepackage[utf8]{inputenc} 
\usepackage[T1]{fontenc}    
\usepackage{hyperref}       
\usepackage{url}            
\usepackage{booktabs}       
\usepackage{amsfonts}       
\usepackage{nicefrac}       
\usepackage{microtype}      
\usepackage{xcolor}         
\usepackage{graphicx}
\usepackage{amssymb}
\usepackage{multirow}
\usepackage[normalem]{ulem}
\useunder{\uline}{\ul}{}
\usepackage{makecell}
\usepackage{enumitem}
\usepackage{wrapfig}

\title{HybridQuestion: Human–AI Collaboration for Identifying High-Impact Research Questions}

\href{}{
\author{
\textbf{Keyu Zhao}$^{1}$\hspace{3mm} 
\textbf{Fengli Xu}$^{1,2*}$\hspace{3mm} 
\textbf{Yong Li}$^{1,2*}$\hspace{3mm} 
\textbf{Tie-Yan Liu}$^{2}$\\
$^1$Department of Electronic Engineering, BNRist, Tsinghua University\\
$^2$Zhongguancun Academy \\
$^*$\{fenglixu, liyong07\}@tsinghua.edu.cn
}
}

\begin{document}

\maketitle

\begin{abstract}
    The "AI Scientist" paradigm is transforming scientific research by automating key stages of the research process, from idea generation to scholarly writing. This shift is expected to accelerate discovery and expand the scope of scientific inquiry. However, a key question remains unclear: can AI scientists identify meaningful research questions? While Large Language Models (LLMs) have been applied successfully to task-specific ideation, their potential to conduct strategic, long-term assessments of past breakthroughs and future questions remains largely unexplored. To address this gap, we explore a human-AI hybrid solution that integrates the scalable data processing capabilities of AI with the value judgment of human experts. Our methodology is structured in three phases. The first phase, AI-Accelerated Information Gathering, leverages AI's advantage in processing vast amounts of literature to generate a hybrid information base. The second phase, Candidate Question Proposing, utilizes this synthesized data to prompt an ensemble of six diverse LLMs to propose an initial candidate pool, filtered via a cross-model voting mechanism. The third phase, Hybrid Question Selection, refines this pool through a multi-stage filtering process that progressively increases human oversight. To validate this system, we conducted an experiment aiming to identify the Top 10 Scientific Breakthroughs of 2025 and the Top 10 Scientific Questions for 2026 across five major disciplines. Our analysis reveals that while AI agents demonstrate high alignment with human experts in recognizing established breakthroughs, they exhibit greater divergence in forecasting prospective questions, suggesting that human judgment remains crucial for evaluating subjective, forward-looking challenges.
\end{abstract}

\newpage
\tableofcontents 

\newpage

\section{Introduction}
\label{introduction}

The advent of the "AI Scientist" paradigm~\cite{gottweis2025towards,weng2025deepscientist,yamada2025ai, schmidgall2025agent} has catalyzed a significant transformation in the methodology of scientific research~\cite{wang2023scientific, lu2024ai}. This ongoing evolution is characterized by the accelerating automation of the complete research lifecycle, spanning from initial literature review~\cite{schmidgall2025agentrxiv, li2025webthinker, team2025tongyi} and idea generation~\cite{yang2024moose, su2025headsbetteroneimproved, wang2024scimon}, through experimental execution~\cite{tang2025ai, seo2025paper2code, lange2025shinkaevolve}, to scholarly writing~\cite{wang2024autosurvey, yan2025surveyforge, ifargan2025autonomous}. This paradigm shift promises to significantly accelerate the pace of discovery and broaden the scope of scientific inquiry.

However, a critical question remains unanswered in this transformation: Can AI scientists identify meaningful research questions? While a growing body of work has successfully employed Large Language Models (LLMs) for task-specific ideation, such as generating specific research ideas~\cite{ghafarollahi2025sciagents, baek2025researchagentiterativeresearchidea} or formulating novel algorithmic approaches~\cite{romera2024mathematical, novikov2025alphaevolve, zhai2025x}, this focus has largely constrained their application to a tactical, problem-solving level. This limitation raises a significant and largely unexplored question: Can the scope of LLM-driven ideation be extended to a strategic level, where the AI is tasked not just with solving a given question, but with identifying what questions matter? More specifically, can LLMs be effectively utilized to identify \textbf{Major Scientific Breakthroughs} within a given period (e.g., 2025) and subsequently to forecast the \textbf{Grand Scientific Questions} that will shape the scientific landscape of the near future (e.g., 2026)?

The joint process of synthesizing past scientific breakthroughs and generating prospective questions, whether executed by LLMs or human experts, inherently yields a large volume of candidate breakthroughs and future grand questions. The magnitude of this resulting candidate pool necessitates a rigorous and efficient methodological framework for prioritization and refinement. Traditionally, the rigorous process of identifying and prioritizing the final list of high-impact scientific contributions or future research directions has relied solely on human-centric evaluation. This established methodology, where human expert panels engage in extensive screening, deliberation, and voting on the large candidate pool, is exemplified by annual reviews from prominent scientific journals or foundation-led priority setting. We seek to explore a novel synthesis of these processes: Can LLMs be effectively integrated into this critical evaluation and filtering stage to create a human-AI hybrid intelligence system?

To address these questions, we designed and implemented a novel three-phase methodology integrating artificial intelligence. The first phase, \textbf{ AI-Accelerated Information Gathering}, is a hybrid process designed to identify high-impact research themes. It begins with a computational analysis of OpenAlex~\cite{priem2022openalex} literature, utilizing keyword embeddings to calculate an annual "hotness" score based on semantic density and frequency. We supplement these quantitative findings with qualitative context from LLM-powered deep research into media and industry discourse. The second phase, \textbf{Candidate Question Proposing}, utilizes this synthesized information to prompt an ensemble of six diverse LLMs (three from the US and three from China) to propose an initial set of 600 candidates. This pool is then refined to 100 candidates via a cross-model voting mechanism to ensure fairness and mitigate model-specific bias. The third and final phase, \textbf{Hybrid Question Selection}, employs a multi-stage human-AI hybrid intelligence framework to refine this pool. This framework progressively escalates human oversight in two stages: it starts with a broad screening phase (100 → 30) with equal human-AI decision weight, followed by an expert-driven refinement phase (30 → 10) employing weighted voting (human:AI = 7:1). This process yields ten finalists, with the two highest-ranked candidates securing their spots in the "Top 10 Major Scientific Breakthroughs of 2025"(or "Top 10 Grand Scientific Questions for 2026").

To define a focused and high-impact scope, we drew inspiration from the disciplinary breadth of influential awards (such as the Nobel Prize~\cite{nobel_copyright_2025}and Turing Award~\cite{acm_turing_award_2025}) to limit our analysis to five key fields: Artificial Intelligence, Physics, Chemistry, Biology, and Economics. To validate this system, we conducted an experiment aiming to identify the Top 10 Major Scientific Breakthroughs of 2025 and the Top 10 Grand Scientific Questions for 2026 across these disciplines. Our results demonstrate that the AI-driven synthesis successfully captured the intrinsic semantic structure of complex fields, generating high-quality candidates that encompass both verified technical milestones—such as pure reinforcement learning systems—and structural open problems. However, a divergence persists based on the nature of the task: while AI agents demonstrate high alignment with human experts in recognizing established breakthroughs, they exhibit greater divergence in forecasting prospective questions. Notably, this divergence is characterized by human experts placing a higher premium on pragmatic, meta-level infrastructure (e.g., evaluation standards), whereas AI agents prioritize theoretical capabilities. This suggests that while AI is a powerful engine for broad-spectrum scanning and initial proposal, human judgment remains crucial for evaluating subjective, forward-looking challenges.

In summary, the main contributions of this work include:
\begin{itemize}[leftmargin=15pt]
    \item We extend the application of LLM-driven ideation from the tactical, problem-solving level to a strategic perspective, demonstrating its capability to conduct comprehensive reviews of past scientific breakthroughs and forecast future grand questions.
    
    \item We design and implement a novel end-to-end methodology that operationally integrates artificial intelligence into both the proposal phase via hybrid data synthesis and the voting phase via a multi-stage human-AI hybrid framework.

    \item We present two curated, high-impact lists generated through our methodology: "Top 10 Scientific Breakthroughs of 2025" and "Top 10 Scientific Questions for 2026". These comprehensive lists serve as strategic compass points, offering researchers and policymakers critical insight into emerging trends and priority research frontiers.
\end{itemize}
\section{Methodology}
\label{section 2}
\subsection{Overview}
\begin{figure}[t]
    \begin{center}
    \vspace{-6mm}
    \includegraphics[width=14cm]{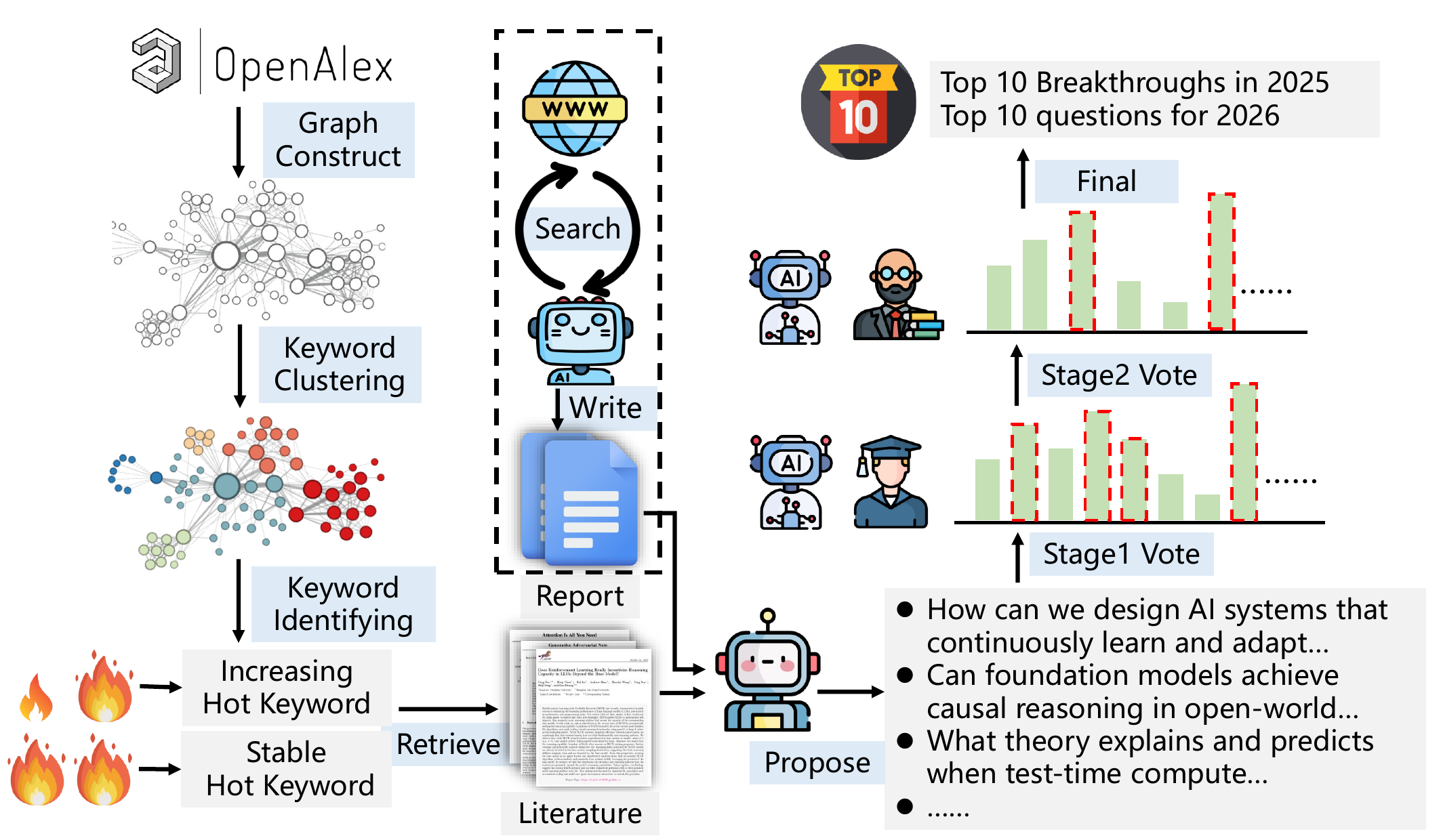}
    \vspace{-6mm}
    \caption{Overview of our proposed human-AI hybrid methodology. }
    \label{figure1}
    \end{center}
\end{figure}

Our methodology is structured into three primary phases: (1) AI-Accelerated Information Gathering, which combines computational literature analysis and web-based deep research; (2) Candidate Question Proposing, an ensemble process to propose and filter a robust candidate pool; and (3) Hybrid Question Selection, a multi-stage framework to screen and finalize the lists. We illustrate the overview of our methodology in Figure ~\ref{figure1}.

\subsection{AI-Accelerated Information Gathering}
\label{section 2.1}
This phase integrates computational analysis of publication data with supplementary web research. By leveraging AI's capability to process vast datasets with unparalleled speed and breadth, we can aggregate and synthesize information from millions of academic papers and diverse media sources far more efficiently than traditional manual review methods. The process is structured into two stages: embedding-based keyword analysis and deep research for supplementary context.

\subsubsection{Embedding-based Keyword Analysis}
\label{section 2.1.1}
The foundation of our approach is a computational analysis of scientific literature sourced from OpenAlex~\cite{priem2022openalex}. We collected all publications and their associated keywords for the period of 2015-2025, processed on a per-year basis.

\textbf{Graph Construction and Embedding:} For each year, we constructed a keyword co-occurrence graph. In this graph, nodes represent keywords, and an edge exists if two keywords co-occur within the same publication. We then employed the node2vec~\cite{grover2016node2vec} algorithm on each annual graph to learn a distinct vector embedding for every keyword for that specific year, providing an annual snapshot of each keyword's semantic context.

\textbf{Hotness Calculation:} We developed a "hotness" score to quantify the annual prominence of each keyword. The score is calculated via a weighted sum: for a specific keyword, we aggregate the appearance frequencies of all other keywords, weighted by a Gaussian kernel function based on cosine distance. This method assigns a higher weight to semantically closer keywords, thereby capturing both the keyword's popularity and its density within the semantic space. The Gaussian kernel's bandwidth is determined dynamically each year; it is set to the distance value found at a pre-defined low percentile (referred to as \texttt{sigma\_perc\_1}) of a large sample from the annual cosine distance distribution.

\textbf{Hotness-Priority Clustering:} To identify thematic clusters, we implemented a hotness-priority greedy clustering algorithm. Keywords are processed in descending order of their hotness score. Each keyword is assigned to the first existing cluster where its cosine distance to the cluster's "seed" (the first, and hottest, keyword in that cluster) is below a \texttt{distance\_threshold}. This \texttt{distance\_threshold} is also determined dynamically each year, set to the distance value found at a different pre-defined percentile (referred to as \texttt{sigma\_perc\_2}) of the same sampled distribution. If no suitable cluster is found and the keyword's hotness rank for the year is within the top \texttt{kw\_hotness\_threshold}, it seeds a new cluster. This results in thematically coherent clusters oriented around high-impact keywords.

\textbf{Keyword Selection:} Based on this temporal analysis, we identified two sets of keywords:
\begin{itemize}
    \item \textbf{Keywords for Major Scientific Breakthrough of 2025:} We selected keywords that demonstrated both established prominence and emerging momentum. The former was defined as having a 2024 hotness rank within the top \texttt{kw\_breakthrough\_threshold}, and the latter as showing an increase in hotness rank in 2025. These selected keywords are referred to as \texttt{breakthrough\_keywords}.
    \item \textbf{Keywords for Grand Scientific Question for 2026:} We first identified high-impact clusters, selecting those whose seed keyword's hotness rank was within the top \texttt{cluster\_hotness\_threshold}. From these top-tier clusters, we extracted two sub-sets of keywords: \texttt{question\_keywords\_1}, representing keywords with high absolute importance (absolute hotness rank within the top \texttt{kw\_question\_threshold}), and \texttt{question\_keywords\_2}, representing keywords with high hotness acceleration (hotness change rank within the top \texttt{kw\_question\_threshold}).
\end{itemize}

\subsubsection{Deep Research for Supplementary Context}
\label{section 2.1.2}
We recognize that significant scientific developments are not exclusively documented in academic literature; they also manifest in media reports, industry advancements, and public discourse. To capture this, we utilized a "Deep Research" approach to supplement the OpenAlex data with broader web intelligence.

\begin{itemize}
    \item \textbf{For Major Scientific Breakthroughs of 2025:} We prompted an LLM to search for "significant challenges widely discussed in 2024 that were substantially solved or advanced in 2025."

    \item \textbf{For Grand Scientific Questions for 2026:} We prompted the LLM to identify "hot, rapidly accelerating, or emerging technological and scientific directions in 2025."
\end{itemize}

To mitigate potential retrieval imbalances or regional biases, we employed two distinct deep research models from United States and China. These models were independently prompted to gather supplementary context. Following the independent generation of context by these two models, we employed a third LLM to systematically integrate and consolidate the two resulting context documents.

\subsection{Candidate Question Proposing}
This phase uses the synthesized data from Section~\ref{section 2.1} to generate and refine the initial candidate pool using a multi-LLM ensemble approach.

\subsubsection{Hybrid Proposing of Initial Candidate Lists}
\label{section 2.2.1}
In the first stage of proposal, we synthesized the information from both the embedding analysis ~\ref{section 2.1.1} and the supplementary context ~\ref{section 2.1.2} to prompt LLMs to generate the candidate lists.

\begin{itemize}
    \item \textbf{Major Scientific Breakthroughs of 2025:} For each \texttt{breakthrough\_keywords}, we retrieved associated high-citation publications published in 2025 from OpenAlex. This academic information was provided to the LLM alongside supplementary context from the deep research, such as media reports and industry advancements. The LLM was tasked with integrating these two sources to summarize and propose an initial list of distinct, high-impact scientific breakthroughs.

    \item \textbf{Grand Scientific Questions for 2026:} A similar hybrid method was used. For each \texttt{question\_keyword\_1} and \texttt{question\_keywords\_2}, we retrieved two sets of literature: (1) high-citation papers published in 2025 (for immediate relevance) and (2) high-citation foundational papers published post-2015 (for long-term context). This literature, combined with the contextual information on emerging trends identified during the deep research, was provided to the LLM. The model was tasked to propose an initial list of significant and valuable scientific questions for 2026.
\end{itemize}

This dual-source approach ensures our initial candidates are grounded in both quantitative academic trends and qualitative, real-world context.

\subsubsection{Ensemble Voting for Bias Mitigation}
To ensure the generation process was fair and mitigated the inherent biases of any single model, we employed a multi-LLM ensemble approach. We selected six large language models, three originating from the United States and three from China, to avoid regional favoritism and ensure a fair, balanced generation process. Each of the six LLMs was independently prompted (using the method from section ~\ref{section 2.2.1}) to generate its own list of 100 candidate breakthroughs and 100 candidate questions. This resulted in an initial, aggregated pool of 600 candidates for each list (100 candidates x 6 models).

To consolidate this diverse pool, we implemented a cross-model voting mechanism. The six LLMs were then tasked to vote on the combined 600-candidate list, with each model casting 100 approval votes. The 100 candidates receiving the highest aggregate vote counts were selected as the final candidate pool for the human-AI hybrid voting phase.

\subsection{Hybrid Question Selection}
Following the proposal of the 100-candidate list from the ensemble, we implemented the final voting phase. This multi-stage, human-AI hybrid intelligence framework is structured to progressively refine the candidate pool through two distinct stages: a broad screening stage (100 to 30) and an expert-driven refinement stage (30 to 10).

\subsubsection{Stage1: Initial Screening, from 100 to 30 candidates}
The first stage served as an initial screening to reduce the 100 candidates to 30. This phase employed \textbf{Approval Voting}, allowing voters to cast one vote for any candidate they deemed viable without limit. The voting panel was composed of 30 human voters at the graduate student level and 70 AI agents simulated to perform at an equivalent level. These 70 AI agents were procedurally generated using a method inspired by Virtual Lab~\cite{swanson2025virtual}. A \textbf{ChairAgent} was manually configured to instantiate these 70 graduate-level voting agents, with each agent assigned a specific \texttt{role}, \texttt{specialization} and \texttt{background}. The generation process was constrained to ensure each AI voter possessed a distinct specialization, maximizing topical coverage, and that the cohort collectively represented a balanced distribution of research backgrounds. In this broad screening phase, human and AI votes were weighted equally, and the 30 candidates with the highest total approval votes advanced to the next stage.

\subsubsection{Stage2: Expert Refinement, from 30 to 10 candidates}
In the second stage, the 30 candidates were further refined to 10 through \textbf{Limited Voting}. The voter panel's expertise was elevated, consisting of 10 human voters at the professor level and 30 AI agents simulated to match this expertise. These 30 professor-level AI agents were generated using the same procedural instantiation method employed in Stage 1, ensuring they were instantiated with appropriate roles, specializations, and backgrounds reflecting this higher level of expertise. Each voter, both human and AI, was allocated exactly 10 votes to distribute among the candidates. To prioritize nuanced expert judgment at this stage, the voting power was recalibrated: each human vote was assigned a weight of 7, whereas each AI vote was assigned a weight of 1. Based on the final tally, the ten highest-scoring candidates are shortlisted, and the top two are inducted into the "Top 10 Major Scientific Breakthroughs of 2025" (or "Top 10 Grand Scientific Questions for 2026").
\section{Results}
\subsection{Implementation Settings}
This section details the specific implementation parameters and models used throughout our three-phase methodology (Section ~\ref{section 2}).

\begin{figure}[!htbp]
    \begin{center}
    \includegraphics[width=14cm]{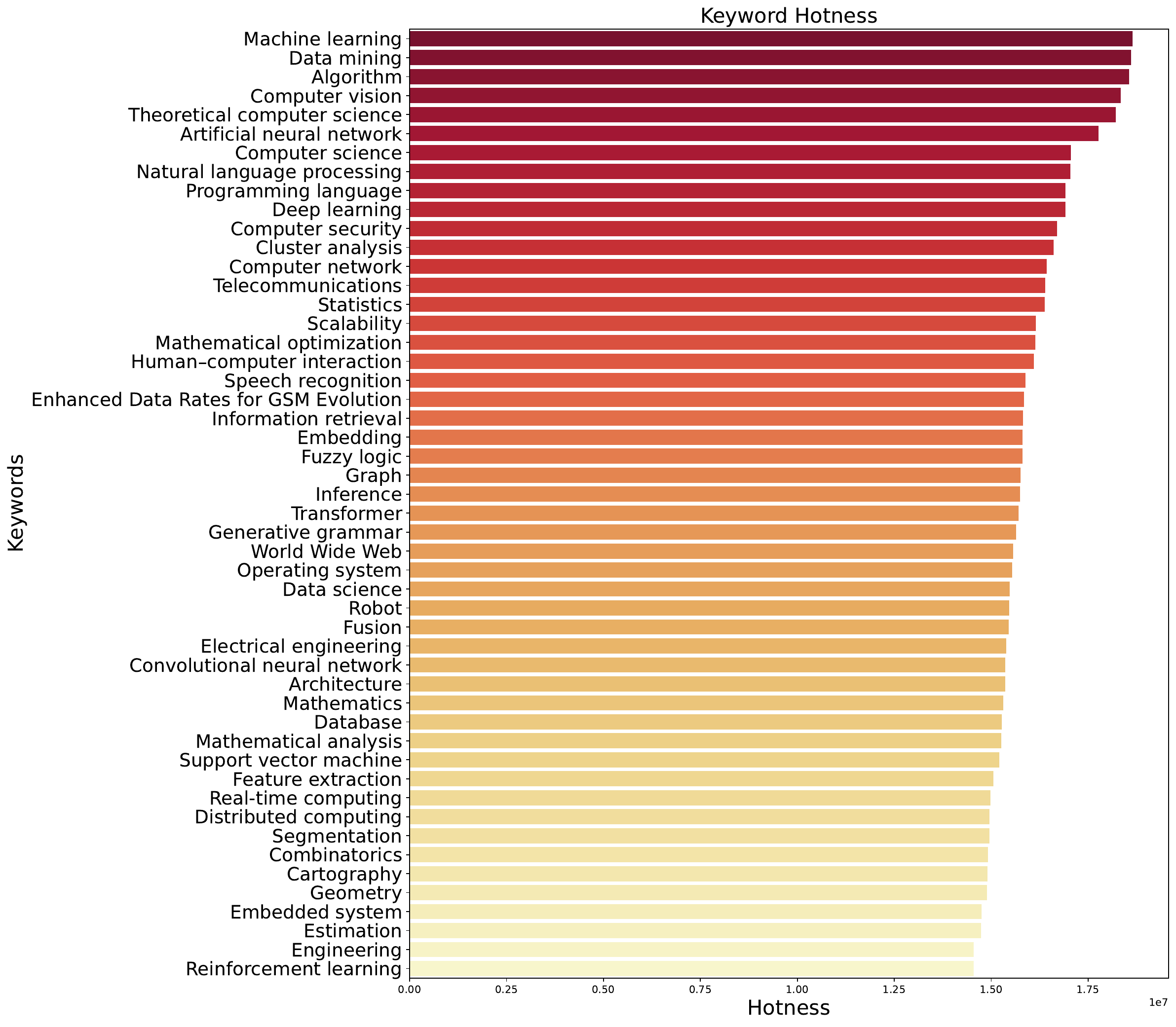}
    \caption{Keyword hotness in the field of Artificial Intelligence in 2025}
    \label{figure2}
    \end{center}
\end{figure}

\subsubsection{General Scope and Data Source}
We first defined our scope, drawing inspiration from the disciplinary breadth of influential awards, such as the Nobel Prize and Turing Award, to focus on five key fields: Artificial Intelligence, Physics, Chemistry, Biology, and Economics. All academic data was sourced from OpenAlex. Disciplinary fields were identified via the "topic" field, and keywords were extracted from the "concepts" field associated with each publication.

\subsubsection{Data Synthesis Settings}
For the embedding-based keyword analysis, we employed the node2vec algorithm. Node2vec is a graph embedding method that learns low-dimensional feature representations for nodes by performing biased random walks on the graph, effectively capturing both local and global network structure. Our implementation used the following parameters: \texttt{embedding\_dim}=128, \texttt{walk\_length}=20, \texttt{num\_walks}=10, \texttt{window\_size}=3, \texttt{p}=1 (return parameter), \texttt{q}=1 (in-out parameter), \texttt{num\_negatives}=5 and \texttt{epochs}=25.

The key threshold parameters were set as follows: \texttt{sigma\_perc\_1} = 0.0005 and \texttt{sigma\_perc\_2} = 0.0005. The hotness-priority clustering used a rank-based percentile threshold of \texttt{kw\_hotness\_threshold} = 0.05. The selection thresholds, which refer to absolute ranks, were set as: \texttt{kw\_breakthrough\_threshold} = 50, \texttt{cluster\_hotness\_threshold} = 5 and \texttt{kw\_question\_threshold} = 5.

\subsubsection{LLM Configuration}
For the Deep Research, we used OpenAI's \texttt{o3-deep-research}~\cite{openai_deep_research_2025} from United States and Qwen's \texttt{qwen-deep-research}~\cite{team2025tongyi} from China. The resulting context documents were subsequently integrated and consolidated using the \texttt{gpt-5}~\cite{openai_gpt5_2025} model.

In the Candidate Question Proposing phase, the ensemble consisted of three US models: \texttt{gpt-5}~\cite{openai_gpt5_2025}, \texttt{gemini-2.5-pro}~\cite{comanici2025gemini}, and \texttt{claude-sonnet-4.5}~\cite{anthropic_claude_sonnet45_2025} and three Chinese models: \texttt{qwen3-max}~\cite{yang2025qwen3}, \texttt{deepseek-v3.1-terminus}~\cite{deepseek_v31_terminus_2025} and \texttt{kimi-k2-thinking}~\cite{team2025kimi}.
For the Human-AI Hybrid Voting phase, \texttt{gpt-5} was used as the backbone model for the ChairAgent that instantiated the voting panel, as well as for all 70 graduate-level and 30 professor-level voting agents.

Across all LLM-based generation and voting tasks, the temperature parameter was set to 0.6.

\subsection{Quantitative Keyword Presentation}
This section visualizes the outcomes of the foundational data synthesis. We utilize the Artificial Intelligence domain as a representative case study for graphical visualizations (Figure ~\ref{figure2}, Figure  ~\ref{figure3}, and Table ~\ref{table1}), while Table ~\ref{table2} and Table ~\ref{table3} detail the keyword selection results across all five studied disciplines.

\subsubsection{Annual Hotness Analysis}
Figure ~\ref{figure2} displays the top 50 keywords in the Artificial Intelligence domain for the year 2025, ranked by their calculated hotness scores. The results highlight the continued dominance of core paradigms, with "Artificial intelligence", "Machine learning" and "Deep learning" occupying the highest ranks. Notably, the list balances methodological pillars (e.g., "Artificial neural network", "Computer vision") with strong theoretical foundations (e.g., "Algorithm", "Statistics"), while also reflecting significant application-oriented trends in areas such as "Natural language processing."

\subsubsection{Semantic Embedding Visualization}

\begin{figure}[!ht]
    \begin{center}
    \includegraphics[width=14cm]{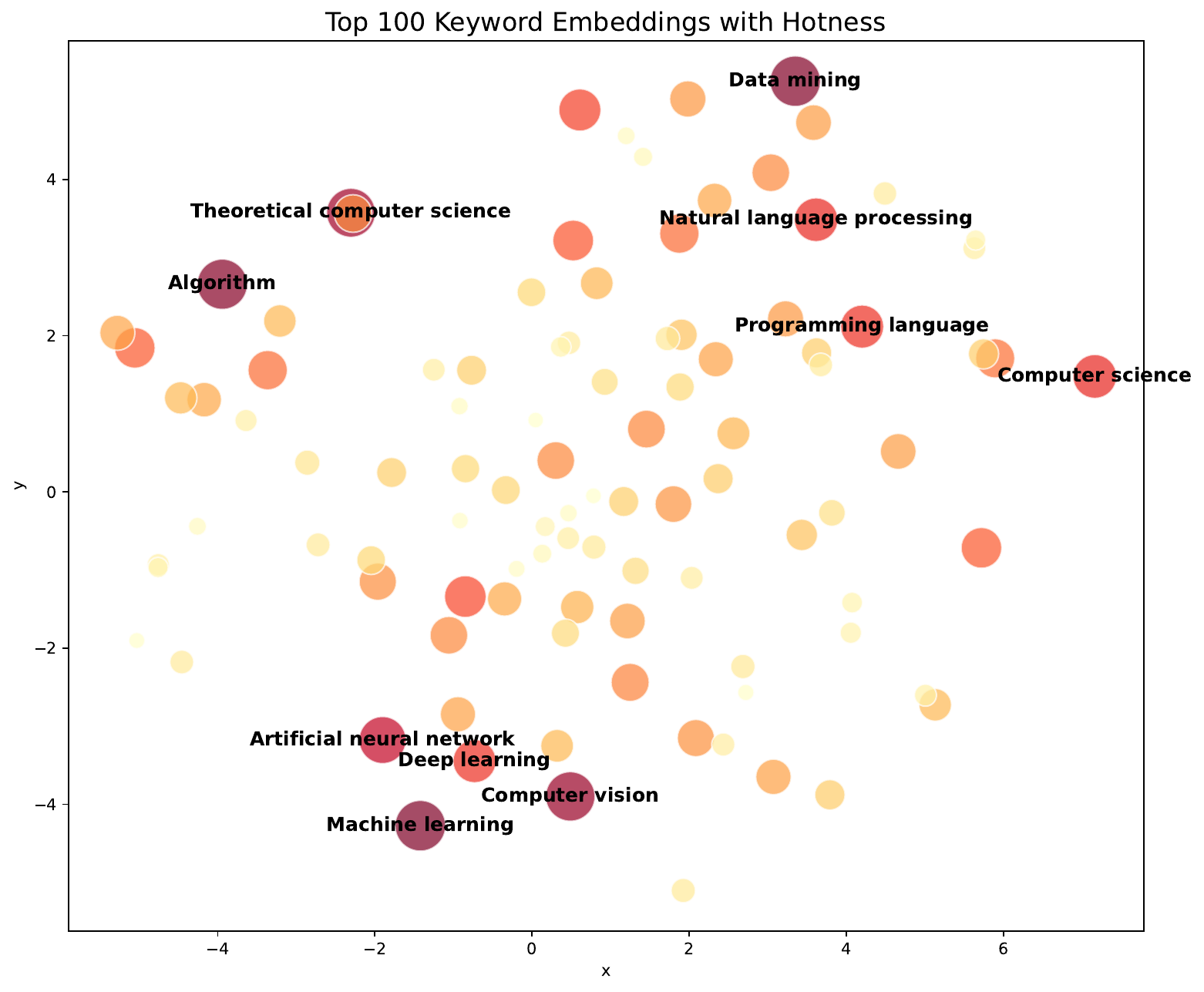}
    \caption{Visualization of keyword embeddings and hotness in Artificial Intelligence in 2025}
    \label{figure3}
    \end{center}
\end{figure}
To provide a structural perspective on the field's topology, we visualized the semantic relationships among the top 100 hottest keywords for the year 2025. Using the 128-dimensional vector embeddings generated via node2vec, we applied t-Distributed Stochastic Neighbor Embedding (t-SNE) to project these high-dimensional representations into a two-dimensional plane. Figure ~\ref{figure3} presents the 2D semantic map of these keyword embeddings. The spatial arrangement reveals clear semantic clustering, effectively grouping related concepts. For instance, neural network-based methods like "Deep learning" and "Artificial neural network" form a tight cluster, distinct from application-focused groups such as "Natural language processing" and theoretical clusters involving "Algorithm". This distribution visually confirms that the embedding model effectively captures the intrinsic semantic structure of the discipline.

\subsubsection{Thematic Clustering and Keyword Selection}
To demonstrate the outcome of our clustering and selection process, we present the organized thematic clusters and the specific keyword sets used for generation.

\textbf{Cluster Presentation:}
\begin{table}[htbp]
\centering
\caption{The Top 5 clusters by hotness and keywords within each cluster}
\label{table1}
\setlength{\tabcolsep}{6pt} 
\renewcommand{\arraystretch}{1.2} 

\begin{tabular}{@{} >{\bfseries}l p{10cm} @{}}
\toprule
\textbf{Cluster} & \textbf{Keyword List} \\
\midrule
Cluster 1 & Machine learning, Data mining, Algorithm, Artificial neural network, Deep learning... \\
\midrule
Cluster 2 & Theoretical computer science, Programming language, Scalability, Embedding, Mathematics... \\
\midrule
Cluster 3 & Computer security, Encryption, Cryptography... \\
\midrule
Cluster 4 & Cluster analysis, Fuzzy clustering, Hierarchical clustering, Spectral clustering\\
\midrule
Cluster 5 & Selection(genetic algorithm), Feature selection \\
\bottomrule
\end{tabular}
\end{table}
Table ~\ref{table1} displays the results of our hotness-priority greedy clustering algorithm (detailed in Section ~\ref{section 2.1.1}) for the Artificial Intelligence domain. The table lists the top 5 thematic clusters, ranked by the hotness of their seed keywords. The results demonstrate high semantic coherence within each group. For instance, Cluster 1 aggregates the central methodological pillars of the field ("Machine learning", "Deep learning", "ANN"), while Cluster 2 effectively isolates the theoretical and foundational aspects ("Theoretical computer science", "Mathematics", "Scalability"). Distinct functional areas are also successfully partitioned, with Cluster 3 focusing on Security ("Encryption", "Cryptography") and Cluster 5 on Optimization ("Genetic algorithms", "Feature selection"). This clustering confirms that our embedding-based approach successfully captures the distinct sub-disciplines within the broader AI landscape.

\textbf{Keyword Selection for Major Scientific Breakthroughs of 2025:}

Table ~\ref{table2} presents the \texttt{Breakthrough\_Keywords} selected across our five target domains. These keywords were identified based on the criteria of having a high hotness rank in 2024 (established prominence) combined with continued growth in 2025. The selection reflects mature, high-impact areas ripe for significant advancement. For example, in Artificial Intelligence, the list focuses on "Machine learning" and "Computer vision," while in Physics, it highlights "Quantum mechanics" and "Optics." In Biology, critical medical frontiers such as "Cancer research" and "Apoptosis" are prioritized. These keywords serve as the "seeds" prompting the LLM to review substantial achievements in these well-defined areas.
\begin{table}[htbp]
\centering
\caption{The selected \texttt{Breakthrough\_Keywords} across five domains}
\label{table2}
\setlength{\tabcolsep}{6pt} 
\renewcommand{\arraystretch}{1.2} 

\begin{tabular}{@{} >{\bfseries}m{1.5cm} m{3.7cm} m{7.5cm} @{}} 
\toprule
\multicolumn{1}{c}{\textbf{Domain}} & \multicolumn{1}{c}{\textbf{Keyword Type}} & \multicolumn{1}{c}{\textbf{Keyword List}} \\ 
\midrule
{\shortstack{Artificial\\Intelligence}} & \texttt{Breakthrough\_Keywords} & Machine learning, Data mining, Feature, Pattern recognition, Computer vision...\\
\midrule
Physics & \texttt{Breakthrough\_Keywords} & Quantum mechanics, Optics, Classical mechanics, Atomic physics, Statistical physics... \\
\midrule
Chemistry & \texttt{Breakthrough\_Keywords} & Catalysis, Physical chemistry, Organic chemistry, Chemical engineering, Inorganic chemistry... \\
\midrule
Biology & \texttt{Breakthrough\_Keywords} & Apoptosis, Cancer research, Downregulation and upregulation, Molecular biology, Inflammation... \\
\midrule
Economics & \texttt{Breakthrough\_Keywords} & Economic growth, Industrial organization, Actuarial science, Natural resource economics, Corporate governance...\\
\bottomrule
\end{tabular}
\end{table}
\textbf{Keyword Selection for Grand Scientific Questions for 2026:}
\begin{table}[htbp]
\centering
\caption{The selected \texttt{Question\_Keywords\_1} and \texttt{Question\_Keywords\_2} across five domains}
\label{table3}
\setlength{\tabcolsep}{8pt}
\renewcommand{\arraystretch}{1.2}

\begin{tabular}{@{} >{\bfseries}m{1.5cm} m{3.7cm} m{7.5cm} @{}}
\toprule
\multicolumn{1}{c}{\textbf{Domain}} & \multicolumn{1}{c}{\textbf{Keyword Type}} & \multicolumn{1}{c}{\textbf{Keyword List}} \\
\midrule
\multirow{2}{*}{\shortstack{Artificial\\Intelligence}} & \texttt{Question\_Keywords\_1} & Machine learning, Data mining, Algorithm... \\
                          & \texttt{Question\_Keywords\_2} & Interest point detection, Natural language processing, Computer vision... \\
\midrule
\multirow{2}{*}{Physics} & \texttt{Question\_Keywords\_1} & Quantum mechanics, Optics, Atomic physics... \\
                          & \texttt{Question\_Keywords\_2} & Thermodynamics, Theoretical physics, Nuclear engineering... \\
\midrule
\multirow{2}{*}{Chemistry} & \texttt{Question\_Keywords\_1} & Catalysis, Physical chemistry, Inorganic chemistry... \\
                          & \texttt{Question\_Keywords\_2} & Hydrogen, Biochemical engineering, Hydrogen bond... \\
\midrule
\multirow{2}{*}{Biology} & \texttt{Question\_Keywords\_1} & Cancer research, Molecular biology, In vitro... \\
                          & \texttt{Question\_Keywords\_2} & Tumor cells, Metastasis, Messenger RNA... \\
\midrule
\multirow{2}{*}{Economics} & \texttt{Question\_Keywords\_1} & Economic growth, Market economy, Industrial organization... \\
                          & \texttt{Question\_Keywords\_2} & Inequality, Socioeconomics, Social science...\\
\bottomrule
\end{tabular}
\end{table}
Table ~\ref{table3} details the question\_keywords used to generate future challenges, categorized into two distinct types as defined in our methodology. \texttt{Question\_Keywords\_1} represents keywords with high absolute hotness, targeting enduring "hard problems" in the field. For instance, in Biology, this includes broad, foundational challenges like "Cancer research" and "Molecular biology." In contrast, \texttt{Question\_Keywords\_2} represents keywords with high acceleration, capturing rapidly emerging trends. In the same Biology domain, this captures more specific, fast-moving topics such as "Tumor cells" and "Messenger RNA". By distinguishing between these two types, our methodology ensures that the forecasted "Grand Questions" address both the foundational unresolved issues of the discipline and its most dynamic, cutting-edge frontiers.

\subsection{Vote Result Presentation}
This section details the progression of our candidate pool through the human-AI hybrid voting framework. We present the initial pool generated by the multi-LLM ensemble and visualize the rigorous filtering process that distills these hundreds of proposals down to the final, high-impact selections.
\subsubsection{Initial Candidate Pool}
\begin{figure}[!ht]
    \begin{center}

    \includegraphics[width=14cm]{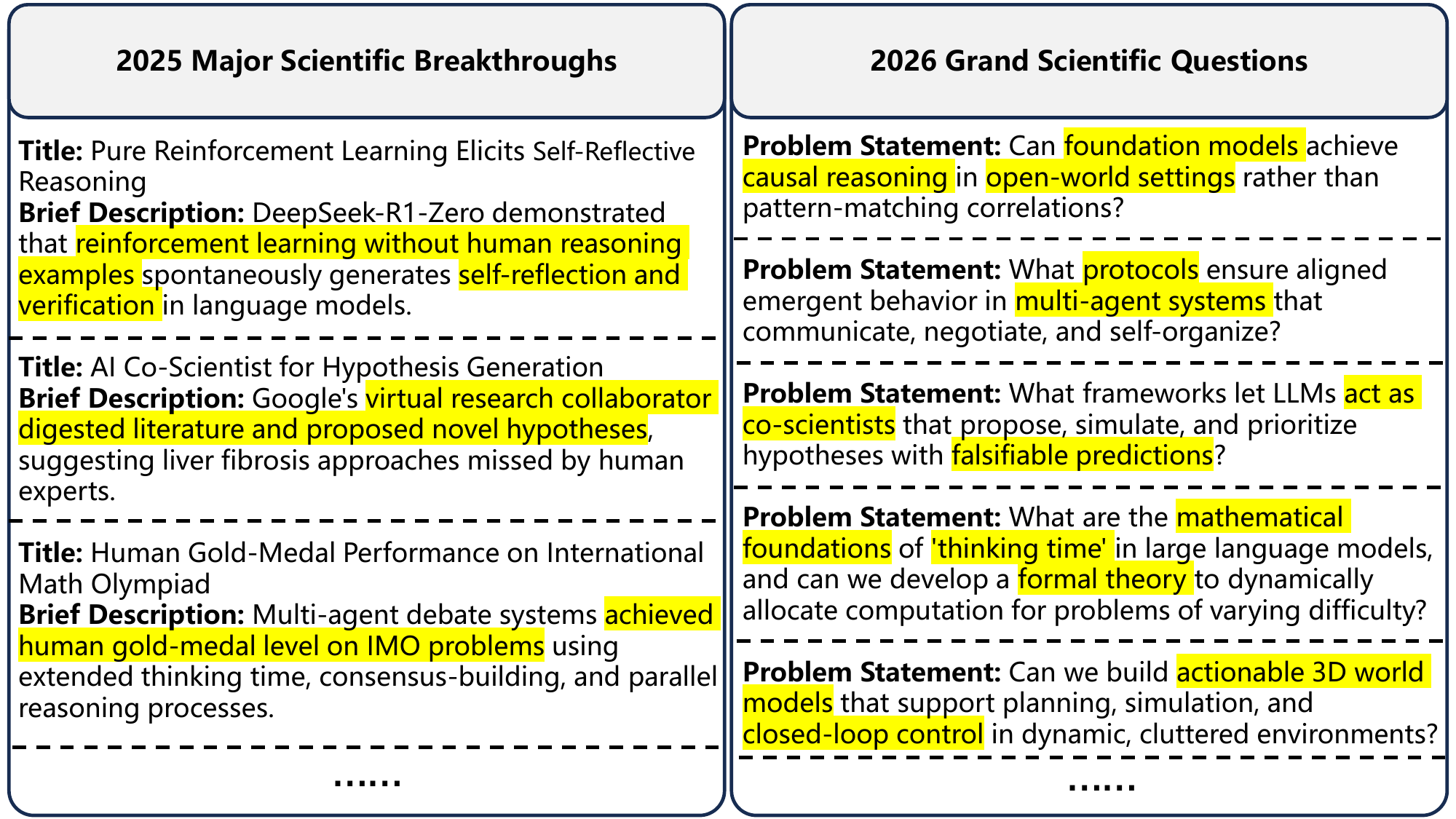}
    \vspace{-6mm}
    \caption{Initial 100 Candidates for the Artificial Intelligence domain}
    \label{figure4}
    \end{center}
\end{figure}

To illustrate the quality and diversity of the proposals generated by our Multi-LLM Ensemble, Figure ~\ref{figure4} presents a representative sample of the "Initial 100 Candidates" for the Artificial Intelligence domain. The figure is divided into two panels to contrast our dual outputs: the left panel, Major Scientific Breakthroughs of 2025, and the right panel, Grand Scientific Questions for 2026. The breakthrough list features specific, retrospective milestones, such as "Pure Reinforcement Learning" (DeepSeek-R1-Zero) and the "AI Co-Scientist" system, emphasizing verifiable technical achievements. In contrast, the question list outlines forward-looking, structural challenges, such as the capability of foundation models to achieve "causal reasoning" in open-world settings or the development of actionable 3D world models for dynamic environments. This sharp distinction demonstrates that our keyword-driven multi-model approach successfully disentangles retrospective validation from prospective inquiry, generating a candidate pool that covers both the consolidated achievements and the critical unresolved frontiers of the discipline.

\subsubsection{A Case of Question Selection}
Figure ~\ref{figure5} maps the selection trajectory for the Grand Scientific Questions in Artificial Intelligence, offering a granular view of candidate evolution. The visualization underscores the rigor of the filtering process by tracing specific proposals from the initial pool (Left) through Stage 1 screening (Middle) to the final selection (Right). Notably, the question addressing "continuous learning in non-stationary environments" consistently garnered high consensus, rising from an initial set of 40 votes to 81 votes in Stage 1, ultimately securing the top position (marked in yellow) in the final stage. Similarly, the inquiry regarding "verifiable reasoning traces" demonstrated resilience, maintaining strong support to become the second winner. In contrast, candidates with lower initial consensus, such as inquiries into "multi-agent protocols" and "scaling laws", were effectively filtered out during the early screening phase. This trajectory confirms that our system successfully amplifies high-quality, consensus-driven signals while discarding less impactful proposals.
\begin{figure}[!ht]
    \begin{center}

    \includegraphics[width=14cm]{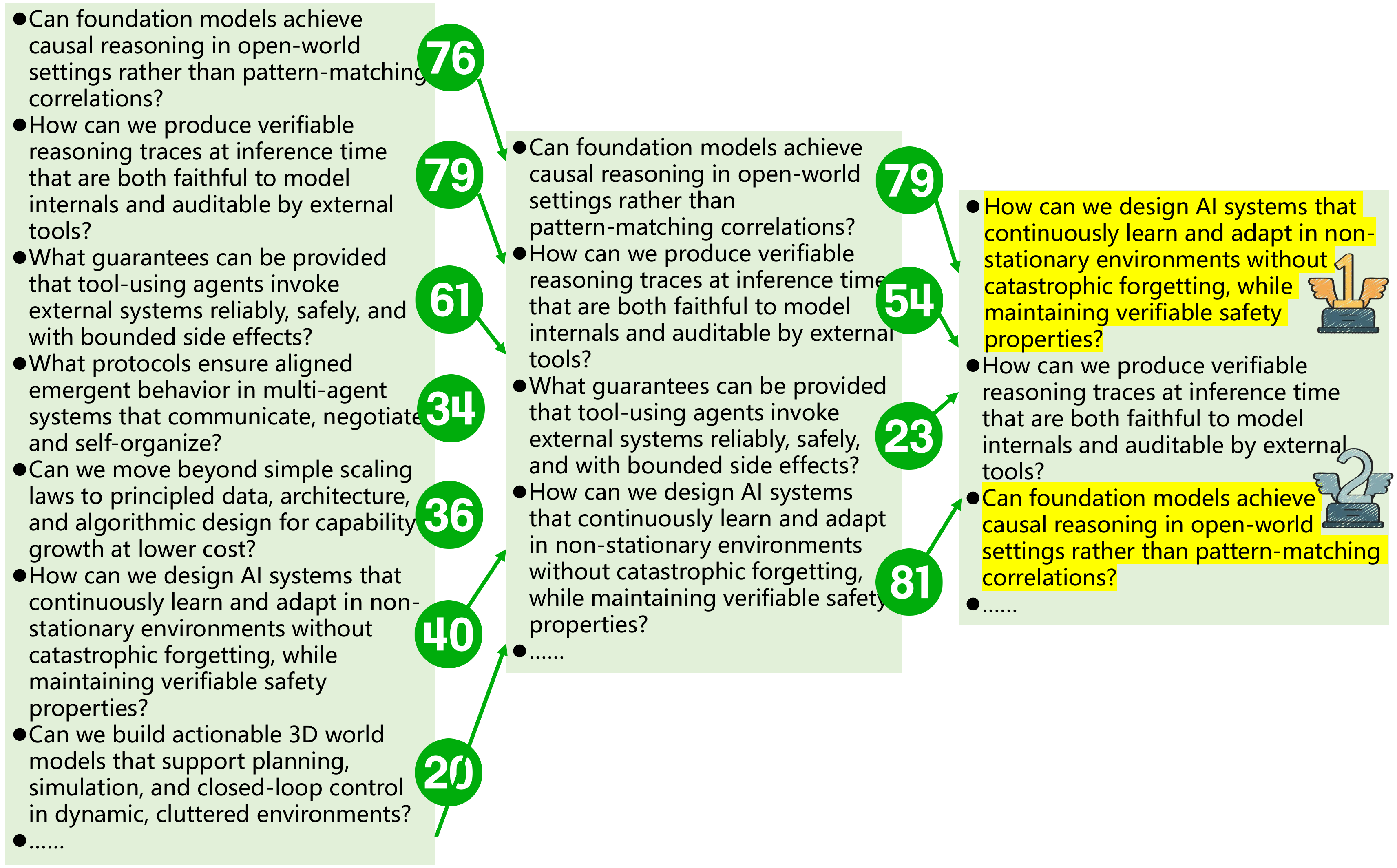}
    \caption{Selection trajectory for the "Grand Scientific Questions" in Artificial Intelligence}
    \label{figure5}
    \end{center}
\end{figure}

\section{Analysis}
To validate the effectiveness of our human-AI hybrid voting framework, we analyzed the alignment between human experts and AI agents across different stages of the voting process. We focus on the Artificial Intelligence domain to examine how voting patterns evolve as the expertise level increases from Stage 1 (Graduate-level) to Stage 2 (Professor-level).
\subsection{Distribution Visualization}
Figure ~\ref{figure5} illustrates the voting distributions for the 30 shortlisted candidates in Stage 2, where the panel consisted of human professors and AI agents simulated to match that expertise. The top row displays the vote counts for the Breakthrough list, while the bottom row displays the Question list. The figure reveals distinct patterns of alignment between human and AI voters.

In the Breakthrough category, the voting distributions are notably similar, particularly in identifying top-tier consensus. For instance, Candidate 1 ("Pure Reinforcement Learning Elicits Self-Reflective Reasoning") received the highest vote count from both human experts and AI agents, indicating a shared recognition of this milestone's significance. This strong alignment suggests that when evaluating established scientific achievements, AI agents effectively mirror human judgment.

Conversely, the Question category exhibits greater divergence. A striking example is Candidate 14 ("How do we build unified, domain‑agnostic evaluation standards that capture real‑world reliability for reasoning, perception, and action?"), which received 7 votes from human experts but failed to garner any votes from AI agents. This discrepancy suggests that human experts may place a higher premium on practical, meta-level infrastructure (like evaluation standards) that ensures reliability, whereas AI agents might prioritize theoretical capabilities, potentially overlooking the critical need for evaluation frameworks. These findings highlight the nuanced differences in how human and AI evaluators prioritize future challenges, with humans showing a stronger inclination towards pragmatic and structural issues.
\begin{figure}[!ht]
    \begin{center}
    \includegraphics[width=14cm]{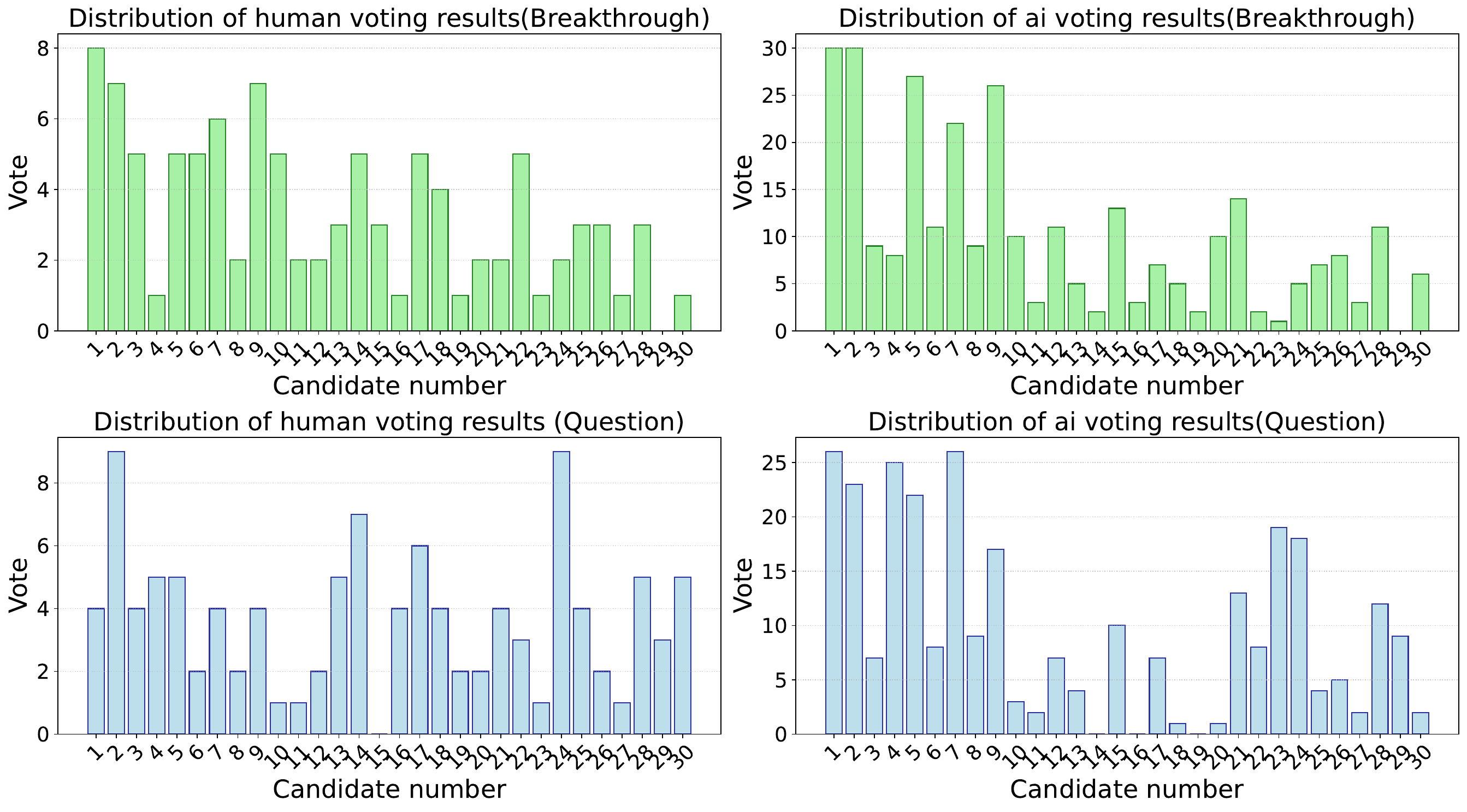}
    \vspace{-6mm}
    \caption{Voting distributions for 30 candidates in Stage 2 in the field of Artificial Intelligence}
    \label{figure5}
    \end{center}
\end{figure}

\subsection{Quantitative Alignment Analysis}
To quantify the alignment between human and AI voting behaviors, we calculated the Jensen-Shannon (JS) Distance between their respective voting distributions. Table ~\ref{table4} presents the JS Distance metric across both Stage 1 and Stage 2 for both the Breakthrough and Question categories.\begin{wraptable}[10]{r}{0.45\textwidth}
    \centering
    \caption{Comparison between human and ai voting distributions}
    \label{table4}
    \begin{tabular}{llc}
    \toprule
    Category      & Phase   & JS Distance \\
    \midrule
    \multirow{2}{*}{Breakthrough} & Stage 1 & 0.394 \\
                                  & Stage 2 & 0.209 \\
    \midrule
    \multirow{2}{*}{Question}     & Stage 1 & 0.370 \\
                                  & Stage 2 & 0.352 \\
    \bottomrule
    \end{tabular}
\end{wraptable}

The results yield two critical observations regarding human-AI alignment. First, there is a notable reduction in JS Distance from Stage 1 to Stage 2 across both categories, with the most significant improvement observed in the Breakthrough list (decreasing from 0.394 to 0.209). This suggests that as the simulated expertise level increases from graduate student to professor, the AI agents' value judgments align more closely with those of human experts. This convergence implies that "Professor-level" AI agents are more effective at capturing the nuanced criteria used by human experts to evaluate high-impact science.

Second, the JS Distance for Breakthroughs is consistently lower than for Questions, particularly in Stage 2 (0.209 vs. 0.352). This disparity likely stems from the inherent nature of the tasks. "Breakthroughs" are grounded in retrospective facts and verifiable impacts, creating a stronger objective ground truth that facilitates consensus. In contrast, "Questions" are prospective and speculative, involving subjective assessments of future value and feasibility. Consequently, even among human experts, there is likely higher variance in opinion regarding future directions, making it inherently more difficult for AI to perfectly mirror human distribution in this category.

Collectively, these quantitative findings affirm that AI scientists are capable of identifying meaningful research questions, albeit with distinct nuances. The high alignment in the retrospective task demonstrates their robust capacity for recognizing meaningful questions answered (Major Breakthroughs in 2025), while the divergence in the prospective task highlights the complexity of forecasting meaningful open questions (Grand Questions in 2026), where human insight remains essential for navigating subjective frontiers.

\section{Conclusion}
In this work, we addressed the critical question of whether AI scientists can identify meaningful research questions. Motivated by the limits of human cognitive bandwidth in the face of exponential literature growth, we proposed a human-AI hybrid framework that combines AI-accelerated data synthesis with a multi-stage collaborative voting mechanism. Through an experimental validation identifying the Top 10 Major Scientific Breakthroughs of 2025 and Top 10 Grand Scientific Questions for 2026, we found that AI agents achieve high alignment with human experts in recognizing retrospective achievements but exhibit notable divergence in forecasting prospective challenges. These findings suggest that while AI scientists can effectively identify meaningful research questions, particularly those grounded in established facts, human judgment remains essential for the subjective, value-driven evaluation of future scientific frontiers.
\bibliographystyle{unsrt}
\bibliography{reference}

\begin{thebibliography}{10}

\bibitem{gottweis2025towards}
Juraj Gottweis, Wei-Hung Weng, Alexander Daryin, Tao Tu, Anil Palepu, Petar Sirkovic, Artiom Myaskovsky, Felix Weissenberger, Keran Rong, Ryutaro Tanno, et~al.
\newblock Towards an ai co-scientist.
\newblock {\em arXiv preprint arXiv:2502.18864}, 2025.

\bibitem{weng2025deepscientist}
Yixuan Weng, Minjun Zhu, Qiujie Xie, Qiyao Sun, Zhen Lin, Sifan Liu, and Yue Zhang.
\newblock Deepscientist: Advancing frontier-pushing scientific findings progressively.
\newblock {\em arXiv preprint arXiv:2509.26603}, 2025.

\bibitem{yamada2025ai}
Yutaro Yamada, Robert~Tjarko Lange, Cong Lu, Shengran Hu, Chris Lu, Jakob Foerster, Jeff Clune, and David Ha.
\newblock The ai scientist-v2: Workshop-level automated scientific discovery via agentic tree search.
\newblock {\em arXiv preprint arXiv:2504.08066}, 2025.

\bibitem{schmidgall2025agent}
Samuel Schmidgall, Yusheng Su, Ze~Wang, Ximeng Sun, Jialian Wu, Xiaodong Yu, Jiang Liu, Zicheng Liu, and Emad Barsoum.
\newblock Agent laboratory: Using llm agents as research assistants.
\newblock {\em arXiv preprint arXiv:2501.04227}, 2025.

\bibitem{wang2023scientific}
Hanchen Wang, Tianfan Fu, Yuanqi Du, Wenhao Gao, Kexin Huang, Ziming Liu, Payal Chandak, Shengchao Liu, Peter Van~Katwyk, Andreea Deac, et~al.
\newblock Scientific discovery in the age of artificial intelligence.
\newblock {\em Nature}, 620(7972):47--60, 2023.

\bibitem{lu2024ai}
Chris Lu, Cong Lu, Robert~Tjarko Lange, Jakob Foerster, Jeff Clune, and David Ha.
\newblock The ai scientist: Towards fully automated open-ended scientific discovery.
\newblock {\em arXiv preprint arXiv:2408.06292}, 2024.

\bibitem{schmidgall2025agentrxiv}
Samuel Schmidgall and Michael Moor.
\newblock Agentrxiv: Towards collaborative autonomous research.
\newblock {\em arXiv preprint arXiv:2503.18102}, 2025.

\bibitem{li2025webthinker}
Xiaoxi Li, Jiajie Jin, Guanting Dong, Hongjin Qian, Yongkang Wu, Ji-Rong Wen, Yutao Zhu, and Zhicheng Dou.
\newblock Webthinker: Empowering large reasoning models with deep research capability.
\newblock {\em arXiv preprint arXiv:2504.21776}, 2025.

\bibitem{team2025tongyi}
Tongyi~DeepResearch Team, Baixuan Li, Bo~Zhang, Dingchu Zhang, Fei Huang, Guangyu Li, Guoxin Chen, Huifeng Yin, Jialong Wu, Jingren Zhou, et~al.
\newblock Tongyi deepresearch technical report.
\newblock {\em arXiv preprint arXiv:2510.24701}, 2025.

\bibitem{yang2024moose}
Zonglin Yang, Wanhao Liu, Ben Gao, Tong Xie, Yuqiang Li, Wanli Ouyang, Soujanya Poria, Erik Cambria, and Dongzhan Zhou.
\newblock Moose-chem: Large language models for rediscovering unseen chemistry scientific hypotheses.
\newblock {\em arXiv preprint arXiv:2410.07076}, 2024.

\bibitem{su2025headsbetteroneimproved}
Haoyang Su, Renqi Chen, Shixiang Tang, Zhenfei Yin, Xinzhe Zheng, Jinzhe Li, Biqing Qi, Qi~Wu, Hui Li, Wanli Ouyang, Philip Torr, Bowen Zhou, and Nanqing Dong.
\newblock Many heads are better than one: Improved scientific idea generation by a llm-based multi-agent system, 2025.

\bibitem{wang2024scimon}
Qingyun Wang, Doug Downey, Heng Ji, and Tom Hope.
\newblock Scimon: Scientific inspiration machines optimized for novelty.
\newblock In {\em Proceedings of the 62nd Annual Meeting of the Association for Computational Linguistics (Volume 1: Long Papers)}, pages 279--299, 2024.

\bibitem{tang2025ai}
Jiabin Tang, Lianghao Xia, Zhonghang Li, and Chao Huang.
\newblock Ai-researcher: Autonomous scientific innovation.
\newblock {\em arXiv preprint arXiv:2505.18705}, 2025.

\bibitem{seo2025paper2code}
Minju Seo, Jinheon Baek, Seongyun Lee, and Sung~Ju Hwang.
\newblock Paper2code: Automating code generation from scientific papers in machine learning.
\newblock {\em arXiv preprint arXiv:2504.17192}, 2025.

\bibitem{lange2025shinkaevolve}
Robert~Tjarko Lange, Yuki Imajuku, and Edoardo Cetin.
\newblock Shinkaevolve: Towards open-ended and sample-efficient program evolution.
\newblock {\em arXiv preprint arXiv:2509.19349}, 2025.

\bibitem{wang2024autosurvey}
Yidong Wang, Qi~Guo, Wenjin Yao, Hongbo Zhang, Xin Zhang, Zhen Wu, Meishan Zhang, Xinyu Dai, Qingsong Wen, Wei Ye, et~al.
\newblock Autosurvey: Large language models can automatically write surveys.
\newblock {\em Advances in neural information processing systems}, 37:115119--115145, 2024.

\bibitem{yan2025surveyforge}
Xiangchao Yan, Shiyang Feng, Jiakang Yuan, Renqiu Xia, Bin Wang, Lei Bai, and Bo~Zhang.
\newblock Surveyforge: On the outline heuristics, memory-driven generation, and multi-dimensional evaluation for automated survey writing.
\newblock In {\em Proceedings of the 63rd Annual Meeting of the Association for Computational Linguistics (Volume 1: Long Papers)}, pages 12444--12465, 2025.

\bibitem{ifargan2025autonomous}
Tal Ifargan, Lukas Hafner, Maor Kern, Ori Alcalay, and Roy Kishony.
\newblock Autonomous llm-driven research—from data to human-verifiable research papers.
\newblock {\em NEJM AI}, 2(1):AIoa2400555, 2025.

\bibitem{ghafarollahi2025sciagents}
Alireza Ghafarollahi and Markus~J Buehler.
\newblock Sciagents: automating scientific discovery through bioinspired multi-agent intelligent graph reasoning.
\newblock {\em Advanced Materials}, 37(22):2413523, 2025.

\bibitem{baek2025researchagentiterativeresearchidea}
Jinheon Baek, Sujay~Kumar Jauhar, Silviu Cucerzan, and Sung~Ju Hwang.
\newblock Researchagent: Iterative research idea generation over scientific literature with large language models, 2025.

\bibitem{romera2024mathematical}
Bernardino Romera-Paredes, Mohammadamin Barekatain, Alexander Novikov, Matej Balog, M~Pawan Kumar, Emilien Dupont, Francisco~JR Ruiz, Jordan~S Ellenberg, Pengming Wang, Omar Fawzi, et~al.
\newblock Mathematical discoveries from program search with large language models.
\newblock {\em Nature}, 625(7995):468--475, 2024.

\bibitem{novikov2025alphaevolve}
Alexander Novikov, Ng{\^a}n V{\~u}, Marvin Eisenberger, Emilien Dupont, Po-Sen Huang, Adam~Zsolt Wagner, Sergey Shirobokov, Borislav Kozlovskii, Francisco~JR Ruiz, Abbas Mehrabian, et~al.
\newblock Alphaevolve: A coding agent for scientific and algorithmic discovery.
\newblock {\em arXiv preprint arXiv:2506.13131}, 2025.

\bibitem{zhai2025x}
Yi~Zhai, Zhiqiang Wei, Ruohan Li, Keyu Pan, Shuo Liu, Lu~Zhang, Jianmin Ji, Wuyang Zhang, Yu~Zhang, and Yanyong Zhang.
\newblock $\backslash$(x$\backslash$)-evolve: Solution space evolution powered by large language models.
\newblock {\em arXiv preprint arXiv:2508.07932}, 2025.

\bibitem{priem2022openalex}
Jason Priem, Heather Piwowar, and Richard Orr.
\newblock Openalex: A fully-open index of scholarly works, authors, venues, institutions, and concepts.
\newblock {\em arXiv preprint arXiv:2205.01833}, 2022.

\bibitem{nobel_copyright_2025}
{Nobel Prize Outreach}.
\newblock Copyright information.
\newblock \url{https://www.nobelprize.org/about/copyright-information/}, 2025.
\newblock Accessed 20 Nov. 2025.

\bibitem{acm_turing_award_2025}
{Association for Computing Machinery}.
\newblock A.m. turing award laureates.
\newblock \url{https://awards.acm.org/turing}, 2025.
\newblock Accessed 20 Nov. 2025.

\bibitem{grover2016node2vec}
Aditya Grover and Jure Leskovec.
\newblock node2vec: Scalable feature learning for networks.
\newblock In {\em Proceedings of the 22nd ACM SIGKDD international conference on Knowledge discovery and data mining}, pages 855--864, 2016.

\bibitem{swanson2025virtual}
Kyle Swanson, Wesley Wu, Nash~L Bulaong, John~E Pak, and James Zou.
\newblock The virtual lab of ai agents designs new sars-cov-2 nanobodies.
\newblock {\em Nature}, pages 1--3, 2025.

\bibitem{openai_deep_research_2025}
{OpenAI}.
\newblock Introducing deep research.
\newblock \url{https://openai.com/index/introducing-deep-research/}, 2025.
\newblock Accessed 20 Nov. 2025.

\bibitem{openai_gpt5_2025}
{OpenAI}.
\newblock Gpt-5.
\newblock \url{https://openai.com/gpt-5/}, 2025.
\newblock Accessed 20 Nov. 2025.

\bibitem{comanici2025gemini}
Gheorghe Comanici, Eric Bieber, Mike Schaekermann, Ice Pasupat, Noveen Sachdeva, Inderjit Dhillon, Marcel Blistein, Ori Ram, Dan Zhang, Evan Rosen, et~al.
\newblock Gemini 2.5: Pushing the frontier with advanced reasoning, multimodality, long context, and next generation agentic capabilities.
\newblock {\em arXiv preprint arXiv:2507.06261}, 2025.

\bibitem{anthropic_claude_sonnet45_2025}
{Anthropic}.
\newblock Claude sonnet 4.5 system card.
\newblock \url{https://www.anthropic.com/claude-sonnet-4-5-system-card}, 2025.
\newblock Accessed 20 Nov. 2025.

\bibitem{yang2025qwen3}
An~Yang, Anfeng Li, Baosong Yang, Beichen Zhang, Binyuan Hui, Bo~Zheng, Bowen Yu, Chang Gao, Chengen Huang, Chenxu Lv, et~al.
\newblock Qwen3 technical report.
\newblock {\em arXiv preprint arXiv:2505.09388}, 2025.

\bibitem{deepseek_v31_terminus_2025}
{DeepSeek AI}.
\newblock Deepseek-v3.1-terminus.
\newblock \url{https://huggingface.co/deepseek-ai/DeepSeek-V3.1-Terminus}, 2025.
\newblock Accessed 20 Nov. 2025.

\bibitem{team2025kimi}
Kimi Team, Yifan Bai, Yiping Bao, Guanduo Chen, Jiahao Chen, Ningxin Chen, Ruijue Chen, Yanru Chen, Yuankun Chen, Yutian Chen, et~al.
\newblock Kimi k2: Open agentic intelligence.
\newblock {\em arXiv preprint arXiv:2507.20534}, 2025.

\end{thebibliography}

\end{document}